# Frictional Dissipation and Scaling Laws at van der Waals Interface: The Key Role of Elastic Pinning of Moiré at Edges and Corners

Xiang Gao,[1,2] Weidong Yan,[3] Wengen Ouyang,[3,4] Ze Liu,[3] Michael Urbakh,[1]*, Oded Hod[1]

[1]Department of Physical Chemistry, School of Chemistry, The Raymond and Beverly Sackler Faculty of Exact Sciences and The Sackler Center for Computational Molecular and Materials Science, Tel Aviv University, Tel Aviv 6997801, Israel

[2]CAS Key Laboratory of Mechanical Behavior and Design of Materials, Department of Modern Mechanics, University of Science and Technology of China, Hefei, 230027, Anhui, China

[3]Department of Engineering Mechanics, School of Civil Engineering, Wuhan University, Wuhan, Hubei 430072, China

[4]State Key Laboratory of Water Resources Engineering and Management, Wuhan University, Wuhan, Hubei, 430072, China

**Abstract**

Van der Waals heterogeneous interfaces are promising candidates for the scaling up of structural superlubricity to meet practical applications. Several factors, however, have been identified that may eliminate superlubricity. Elasticity is one such intrinsic factor, where shear induced lattice reconstruction leads to local interfacial pinning, even at clean pristine contacts. Here, through detailed atomistic simulations, we reveal that incomplete moiré tile pinning at the corners and edges of finite sliders dominates friction from the nano- to the microscales. We further demonstrate that slider shape tailoring and twisting allow to control energy dissipation and its scaling with contact size, thus opening the way to achieve large-scale superlubricity.

*Corresponding author: Michael Urbakh
Email: urbakh@tauex.tau.ac.il

**Introduction**

Structural superlubricity (SSL), the intriguing phenomenon of ultralow friction and wear occurring at incommensurate solid-solid crystalline interfaces, provides vast opportunities for the reduction of energy dissipation and wear in mechanical systems at various length scales.[1] This unique phenomenon was first demonstrated experimentally for twisted graphitic nano-contacts and later extended to incommensurate microscale junctions.[2,3] However, for such homogeneous contacts, SSL is prone to dynamical reorientation of the slider into the commensurate low energy and high-friction configuration.[4] Conversely, van der Waals (vdW) heterostructures, such as graphene (Gr)/hexagonal boron nitride (*h*-BN), Gr/molybdenum disulfide (MoS$_2$), and *h*-BN/MoS$_2$ present an intrinsic lattice mismatch (e.g., ~1.8% for Gr/*h*-BN, ~26.8% for Gr/MoS$_2$, and ~24.6% for *h*-BN/MoS$_2$), such that even the aligned configuration remains incommensurate. This, in turn, prevents interlocking of the interface into a high friction state, leading to superlubric behavior that is robust against interfacial rotations.[5,6] This effect, which was first predicted theoretically,[5] was recently demonstrated experimentally for microscale contacts under ambient conditions.[7,8]

The scaling-up of robust SSL, however, results in the emergence of new energy dissipation mechanisms including interfacial elasticity effects,[9-12] grain boundary dynamics,[13-17] surface, edge and lattice defect reactivity,[18-21] and contaminant intercalation.[22-24] The former, is often associated with bulk frictional contributions, which are expected to appear at the micro- and macro-scales.[25,26] Nonetheless, edge elasticity effects may kick-in already at smaller contact dimensions. The study of such important factors, however, is often limited by the computational burden involved in simulating the dynamics of large-scale junction models. Hence, it is desirable to formulate general scaling laws that can extrapolate information gained in smaller-scale junction model towards the understanding of macroscale elasticity effects.

While experiments on layered material contact sliding measure their dynamical friction properties that inherently involve edge and surface elasticity effects,[8,18,19,27-31] previous computational studies focused mainly on the static friction characteristics of rigid model systems.[32-34] To bridge this gap in this paper, we present the results of large-scale (up to hundreds of nanometers in size) fully atomic simulations of flexible multilayer Gr/*h*-BN interfaces of various contact geometries. We find that elastic pinning of incomplete moiré tiles at edges and corners of the slider plays a key role in energy dissipation in aligned and marginally twisted layered interfaces. This leads to distinct periodicity and shape dependence of the kinetic friction scaling laws. We further identify a sublinear-to-linear transition of the kinetic friction scaling with contact size when surface moiré motions become significant.

**Simulation Setup**

Our model system for Gr/$h$-BN heterojunctions, shown in Fig. 1a, contains a three-layer-thick finite sized ABA- (Bernal-) stacked Gr flake, hydrogen-saturated at its rim atoms, residing atop a three-layer laterally periodic AA'A-stacked $h$-BN substrate. In the aligned configuration the interface exhibits maximal moiré superstructure dimensions with a period of $\lambda_m \approx 13.9$ nm. As shown in Fig. 1b, the top Gr layer is set to be rigid and driven along the armchair ($x$) direction of the substrate with a constant velocity of $v_0 = 5$ m/s. The bottom $h$-BN layer is frozen at its initial position. The intralayer interactions are described by the REBO potential[35] for graphene and the Tersoff potential[36] for $h$-BN, respectively. The interlayer interactions are accounted for by the registry-dependent interlayer potential (ILP).[37-42] To evaluate the effect of contact geometry, we consider four regular polygonal flake shapes, i.e., square, rectangular, triangular, and hexagonal. The scanline position is chosen such the geometric center of the Gr flake slides along a crystallographic axis of the substrate, such that it crosses B and N atoms along its path. In this way, the emerging moiré superstructures present mirror symmetry with respect to the line crossing the center of the flake parallel to the $x$-axis (see Fig. 1c). Notably, due to the finite size of the flake, incomplete moiré tiles appear at flake edges.

The sliding simulations are performed at zero temperature using damped dynamics as done in previous studies.[13,14] Here, to remove the heat generated during shear, viscous damping is applied to the (relative) velocities of the atoms in the second Gr layer ($l_2$) and the second $h$-BN layer ($l_5$) with a damping coefficient $\eta = 1.0$ ps$^{-1}$ in all three directions. This allows heat to dissipate via both the slider and the substrate aiming to mimic experimental scenarios, while introducing minor disturbance to the dynamics of atoms at the shear interface. All simulations are conducted in the LAMMPS package[43] (see Section (Sec.) 1 of the Supporting Material (SM) for further simulation details).

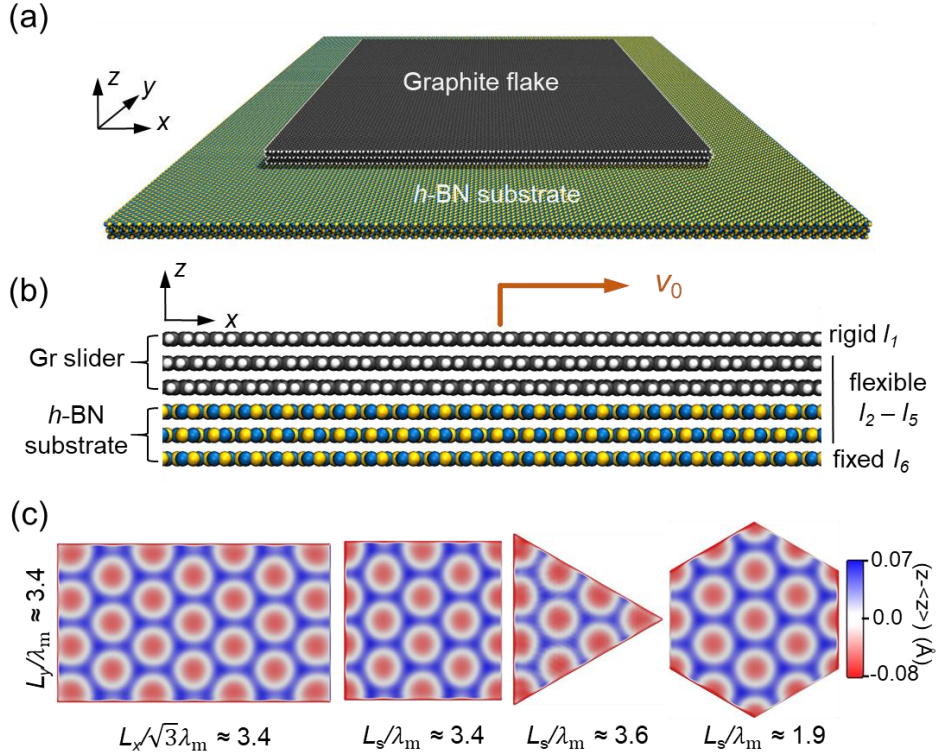

**Figure 1. Model system setup.** (a) Perspective view of the Gr/$h$-BN heterostructure. From top to bottom, the system consists of a three-layer thick Gr slider with hydrogen-saturated rim atoms (black and white spheres, respectively) atop a three-layer periodic $h$-BN substrate (blue and yellow spheres, respectively). (b) Side view of the Gr/$h$-BN heterostructure. (c) Representative color maps of the topographies of the interfacial relaxed Gr layer ($l_3$) for four geometries signifying the emerging moiré superstructures. In the scale bar, $z$ represents the atomic height and $\langle z \rangle$ is the average height. $L_x$, $L_y$ and $L_s$ denote $x$- and $y$-axis side length of the rectangular flake and the side length of all other regular polygons, respectively, and $\lambda_m \approx 13.9$ nm is the moiré period in the aligned configuration.

**Static and kinetic friction dependence on the dimensions of aligned polygonal flakes**

We start by examining rectangular flakes, which allow to independently study the frictional dependence on the two lateral flake dimensions, $L_x$ and $L_y$. The flakes are positioned in the aligned configuration, such that two of their parallel edges are perpendicular to the sliding ($x$) direction and parallel to one of the moiré supercell lattice vectors. Fig. 2a-b show the static (panel a) and the kinetic (panel b) friction forces as a function of $L_x$ (black circles) or $L_y$ (red diamonds), where the other lateral flake dimension remains fixed. Interestingly, the static and kinetic friction forces, $F_s$ and $F_k$, demonstrate distinct behaviors. When increasing $L_x$, the static friction demonstrates pronounced oscillations about a constant average, whereas the kinetic friction remains nearly independent of the edge size for the contact dimensions considered. Conversely, with growing $L_y$, the static friction increases linearly while its kinetic counterpart exhibits strong oscillations with some apparent overall growing envelope. Notably, the periodicities of the static friction with $L_x$ and the kinetic friction with $L_y$ are determined by the moiré superstructure periodicity along the corresponding direction, such that $\lambda_s^{(x)} \sim \sqrt{3}\lambda_m/2$ and $\lambda_k^{(y)} \sim \lambda_m$, respectively. This suggests that the superstructure play an important role in the diverse size dependencies obtained. (See SM Sec. 2 for details of friction calculations.)

When keeping the aspect ratio of the rectangular (or square) flake constant, a linear combination of the individual side dependencies is expected to occur. This can be clearly seen in Fig. 2c, d where the static and kinetic friction forces exhibit oscillations with a linearly growing envelope and periods of $\lambda_s \sim \sqrt{3}\lambda_m/2$ and $\lambda_k \sim \lambda_m$, respectively, for both the rectangular and square flakes when the contact encompasses at least a single moiré supercell. For smaller contact dimensions, edge and corner pinning of the lower graphene layer to the underlying *h*-BN surface result in increased kinetic friction (see Fig. 2d) and may induce stacking instability (ABA⇌ABC) or permanent stacking transformation (ABA⇒ABC) in the tri-layer Gr flake.[6] We note that the positions of extremum slightly vary with flake thickness for $F_k$ but not for $F_s$ (see SM Sec. 3), suggesting that the thickness of flake mainly has a quantitative effect on $F_k$ and $F_s$, while does not qualitatively affect their distinct periodic behaviors.

A similar behavior is obtained for the triangular flakes, where both the static and the kinetic friction forces exhibit linear scaling with side length and oscillations of period $\lambda_m$ (See SM Sec. 4). Notably, while the qualitative behavior of kinetic friction force of the hexagonal flake is the same, the static friction force shows oscillations of half the period $0.5\lambda_m$ with side length.

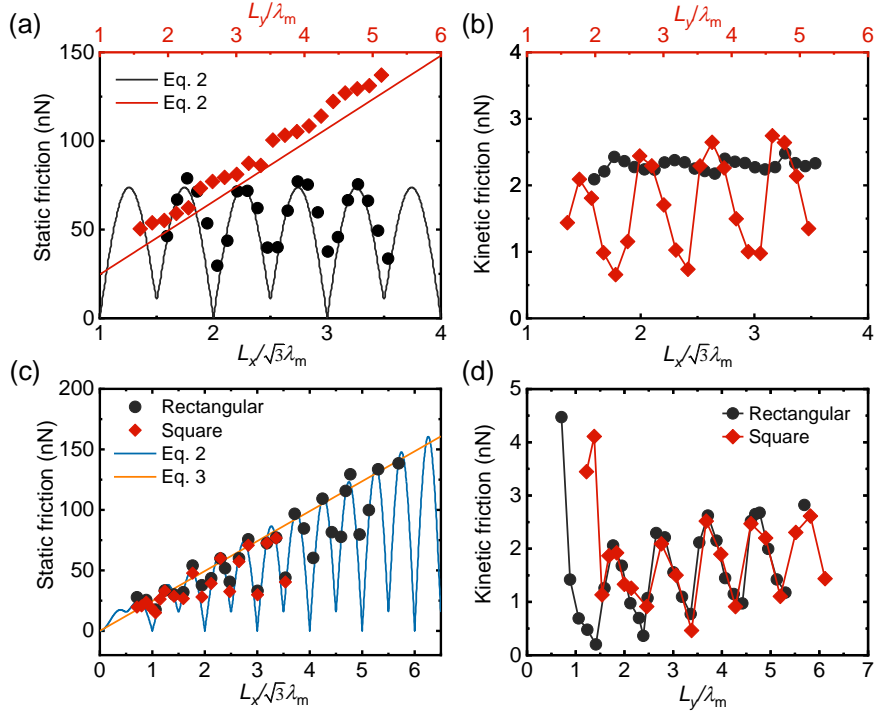

**Figure 2. Scaling of the static and kinetic friction forces with the dimensions of rectangular and square flakes.** (a) $F_s$ and (b) $F_k$ for rectangular flakes as a function of side length. Black symbols (bottom horizontal axis) and red symbols (upper horizontal axis) correspond to the conditions of a fixed $L_y$ ($L_y/\lambda_m = 2.76$) and a fixed $L_x$ ($L_x/\sqrt{3}\lambda_m = 2.76$), respectively. (c)-(d) present $F_s$ and $F_k$, respectively, as a function of the dimensions of square flakes (red) and rectangular flakes (black) with a fixed aspect ratio of $L_x/L_y = \sqrt{3}$. In panel (c), for better comparison, the values of $F_s$ for square flakes are reduced by a factor of $\sqrt{3}$. The solid lines in panels (a) and (c) are theoretical predictions obtained using Eqs. (2) and (3).

**Analytical model for static friction**

The static friction variation with flake dimensions obtained via the atomistic simulations can be rationalized by an analytical model that is based on rigid flake sliding considerations. For the graphene/$h$-BN interface, the interlayer interactions can be approximated by a continuum potential energy density function of the form:[34,44]

$$U(x,y) = -\frac{2}{9}U_0 \left(2\cos\frac{2\pi x}{\sqrt{3}\lambda_m}\cos\frac{2\pi y}{\lambda_m} + \cos\frac{4\pi x}{\sqrt{3}\lambda_m}\right), \quad (1)$$

where $U_0 = 4.5$ meV/Å$^2$ is energy corrugation parameter, consistent with high accuracy density functional theory (DFT) calculation reference.[45] This coarse-grained potential map describes moiré supercell scale interactions and does not account for atomistic features. The total interlayer energy of a flake of surface area $A$ can be calculated as $E(A, x_0, y_0) = \int_{A_{(x_0,y_0)}} U dA$, where $(x_0, y_0)$ is the geometric center of the flake with respect to the underlying substrate. The corresponding static friction can be evaluated as the maximal derivative of the interlayer energy along the scanline (chosen here to be parallel to the $x$ axis): $F_{s,\text{rigid}} = \left(\frac{\lambda_m}{a_{h\text{BN}}}\right) \max_{x_0}\left(\frac{dE}{dx_0}\right)$, where $a_{h\text{BN}} = 2.5045$ Å is the lattice period of $h$-BN. The pre-factor accounts for the fact that in the coarse-grained potential, all distances are scaled with respect to the moiré super-cell dimensions, whereas the derivatives should be taken with respect to the atomic lattice unit-cell displacements. We note that this analytical model, based on the rigid flake assumption, was shown to provide excellent agreement with atomistic force-field calculations of the static friction of rigid layered contacts.[34]

Using this analytical scheme, the static friction of rectangular shaped flakes with $L_y \gg \lambda_m$, can be approximated as (see SM Sec. 5):

$$F_{s,\text{rigid}} \approx \frac{4}{9}\frac{\lambda_m L_y U_0}{a_{h\text{BN}}} \left|\sin\frac{2\pi L_x}{\sqrt{3}\lambda_m}\right|, \qquad (2)$$

and its upper envelope reads:

$$F_{s,\text{rigid}}^{\text{up}} = \frac{4}{9}\frac{\lambda_m L_y U_0}{a_{h\text{BN}}}. \qquad (3)$$

Eqs. (2) and (3) agree well with the MD simulation results for flexible interfaces presented in Fig. 2a and 2c fully accounting for the linear scaling with $L_y$ and the periodic oscillations with $L_x$.

The good agreement between the atomistic simulation results and the coarse-grained analytical model predictions, allows us to identify the physical origin of the different features appearing in the force scaling curves. First, when integrating over complete moiré supercells the contribution of the periodic functions appearing in the potential function of Eq. 1 vanishes. As such, the main contribution to the sliding energy barrier, and hence to the static friction, is attributed to incomplete moiré tiles near the flake edges. Specifically, the maxima and minima of the static friction force as a function of flake dimensions correspond to in-phase (constructive) and opposite-phase (destructive) evolution of incomplete moiré tiles at different edges. For aligned rectangular flakes this effect can be readily demonstrated. On each of the sides that align with the moiré lattice direction ($y$), incomplete edge moiré tiles evolve in phase during sliding along the $x$ direction (see SM Movie 1), leading to constructive contribution to the sliding energy barrier and linear growth of the friction with the $L_y$ side length. On the two sides that are parallel to the sliding direction, the exit of edge moiré tiles during sliding is compensated by the entrance of new moiré tiles on the same side, resulting in no overall growth of the energy barrier with $L_x$. Nonetheless, variation of $L_x$ controls the relative phase of incomplete moiré tile evolution along the two $L_y$ sides, resulting in static friction oscillations of period $\lambda_s^{(x)} \sim \sqrt{3}\lambda_m/2$. We note that while the rigid model predicts zero friction at the oscillation minima, due to perfect compensation of incomplete moiré tiles on different edges,[34] the atomistic simulations demonstrate finite friction with non-zero scaling of the lower oscillation envelope. This is mainly attributed to edge elasticity effects, which are not accounted for in rigid sliding calculations. A similar behavior is seen also for triangular and hexagonal shaped flakes (see SM Sec. 4).

**Energy Dissipation Analysis**

Similar to static friction, kinetic friction in our system is also expected to be dominated by edge effects. To demonstrate this, we performed dynamic sliding simulations and analyzed the spatial distribution of energy dissipation channels across the flake. We find that heat is mainly generated at the corners of the bottom graphene flake, where local in-plane vibrational modes are excited during sliding (see SM Fig. S8). This heat propagates via interlayer interactions to the middle graphene layer, where it is dissipated using the damping scheme described in the Methods section, allowing us to evaluate the dissipated power as demonstrated in Fig. 3. The total energy dissipation is found mainly accomplished by the in-plane (mainly sliding direction) motions and dominated by the contribution from the damped Gr flake layer $l_2$ (see SM Sec. 6). The dominance of corner energy dissipation is attributed to the higher flexibility of atoms in these regions that allows for forming local lattice commensurability with the adjacent layers. In aligned or marginally twisted heterogeneous interfaces, in addition to rim atom pinning effects,[8,20] this leads to incomplete moiré tile pinning, which dominates friction (see SM Movie 2). Notably, we find that the sharper the corner is the higher the energy dissipation power is, such that equilateral hexagonal flakes dissipate less than square and rectangular flakes which, in turn, dissipate less than equilateral triangular flakes with similar dimensions. In particular, for triangular flakes, this effect induces local corner stacking instability (ABA⇌ABC) during sliding (see SM Movie 3), which results in enhanced energy dissipation compared to other polygonal flakes.

The length, $L_y$, of the polygon side that is perpendicular to the sliding direction dictates the pattern evolution of the incomplete moiré tiles at the corners (see Fig. 1c) in sliding, and therefore defines the periodicity $\lambda_k^{(y)} = \lambda_m$ of the kinetic friction. Furthermore, residual energy dissipation appearing mainly along the perpendicular sides (see, e.g., Fig. 3d) is responsible for the linear growth of friction with flake dimensions observed in, e.g., Fig. 2d. We note that since the contribution of the inner flake surface and of polygon sides that lie parallel to the sliding direction is minor, square and rectangular shaped flakes demonstrate similar kinetic friction scaling behavior with system dimensions.

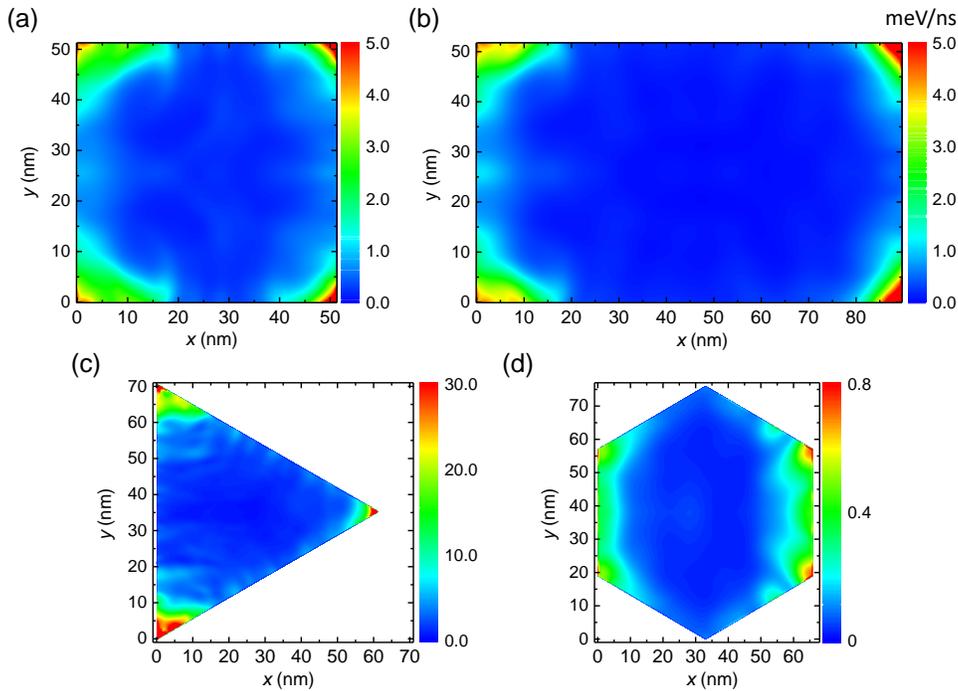

**Figure 3. Energy dissipation power distribution for different flake geometries.** (a)-(d) Energy dissipation maps of the middle layer of (a) square, (b) rectangular (with aspect ratio of $\sqrt{3}$), (c) triangular, and (d) hexagonal graphene flakes measured for system dimensions that correspond to local maxima in the kinetic friction with side length. Atomic energy dissipation power is calculated by averaging over four sliding periods.

**Frictional Scaling Laws**

The mechanism revealed above for corner and edge energy dissipation allows us to devise general friction scaling laws with contact area for flexible layered heterojunctions. Table 1 presents the scaling of the upper envelope of the static friction oscillations for various flake shapes obtained using the analytical rigid sliding model presented above (see Eq. 3 above and SM Sec. 5). As shown in Fig. 4a-c, these expressions provide excellent agreement with MD simulation results. Notably, the static friction scaling is dominated by flake side contribution that is proportional to the square root of the contact area. In this case, the corner contributions are negligible and so are those of the inner surface area.

**Table 1. Analytical expressions for the upper envelopes of different shapes**

| Shape | Upper envelope $F_{s,\text{rigid}}^{\text{up}}(A)$ |
|---|---|
| Square | $\frac{4}{9}\frac{\lambda_m U_0}{a_{h\text{BN}}}A^{\frac{1}{2}}$ |
| Rectangular | $\frac{4}{9\sqrt[4]{3}}\frac{\lambda_m U_0}{a_{h\text{BN}}}A^{\frac{1}{2}}$ |
| Hexagonal | $\frac{(\sqrt{3}+\sqrt{11})\sqrt{5\sqrt{3}+\sqrt{11}}}{36}\frac{\lambda_m U_0}{a_{h\text{BN}}}A^{\frac{1}{2}}$ |
| Triangular | $\frac{8}{9\sqrt[4]{3}}\frac{\lambda_m U_0}{a_{h\text{BN}}}A^{\frac{1}{2}}$ |

For the kinetic friction, an analytical treatment, similar to the rigid model for static friction, is lacking. Nonetheless, considering the three main contributions for kinetic friction, namely: (i) corners; (ii) edges; and (iii) internal surface, we can suggest the following phenomenological scaling law:

$$F_k^{\text{up}}(A) = a_k + b_k A^{\frac{1}{2}} + c_k A, \tag{4}$$

where $a_k$ represents the constant corner contribution, $b_k$ represents the side contribution which scales as the square root of the flake area, and $c_k$ represents the inner surface atomic scale and moiré friction components that scale linearly with area.[9,11] Since the corner and edge contributions are expected to dominate over the surface counterpart in the studied heterogeneous interface, the last term in Eq. (4) can be neglected for the flake dimensions considered in the atomistic MD simulations. The good fit obtained for this expression with the upper (and lower) envelopes of the kinetic friction dependence on contact area for all flake shapes considered (see thick (and thin) solid lines in Fig. 4e-g) supports this proposition. In sharp contrast to the static friction case, the fits to the kinetic friction results yield a dominant contribution to the corner dynamics over all other friction components for the flake dimensions considered (up to $10^4$ nm$^2$ in area). The fitted $a_k$ and $b_k$ values obtained for the different flake shapes follow an order dictated by sharpness of the corner angle, consistent with the picture that emerged above for the energy dissipation maps.

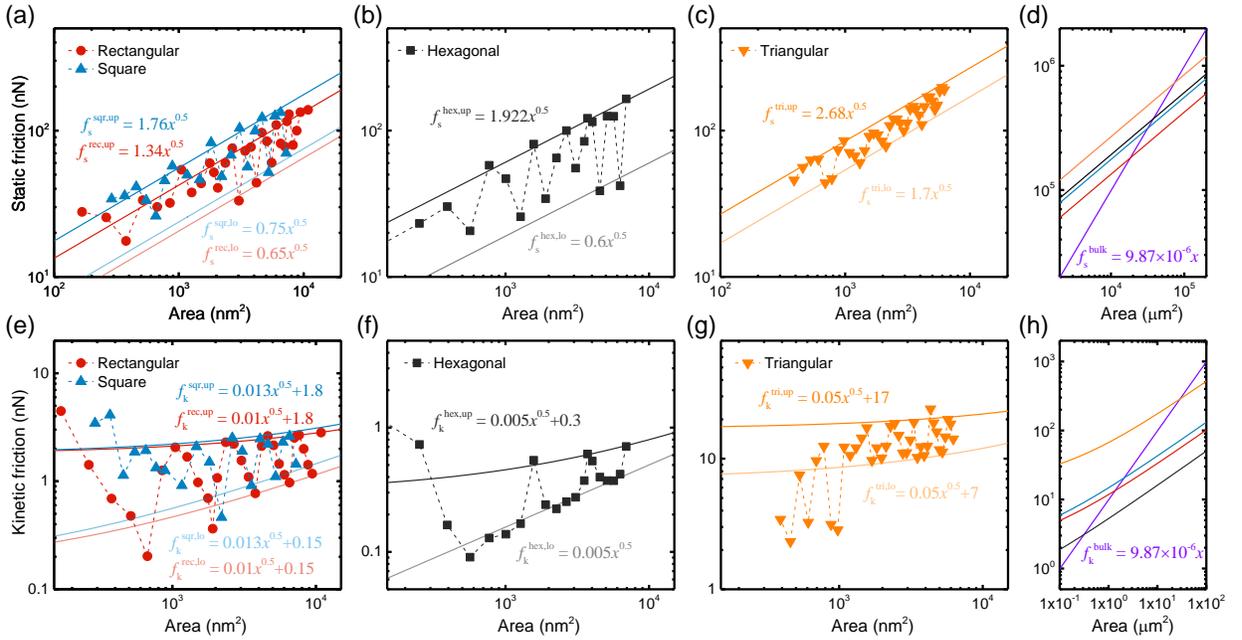

**Figure 4. Static and kinetic frictional scaling with contact dimensions.** (a)-(c) Static friction scaling for (a) rectangular and square; (b) hexagonal; and (c) triangular shaped graphene flakes sliding on an $h$-BN substrate. (d) Extrapolation to large contact area demonstrating the cross-over between corner, edge, and surface dominated friction. Solid symbols connected by dashed lines are atomistic MD simulation results. Dark solid lines (colored according to the corresponding symbol) are calculated using Table 1, with the value of $U_0 = 4.5$ meV/Å$^2$ close to high accuracy DFT calculation reference[45] for the upper envelope. Light solid lines are fits of the lower envelopes to a scaling law proportional to the side length, $F_s \propto A^{\frac{1}{2}}$. (e)-(h) same as (a)-(d) for kinetic friction, respectively, with the solid lines fit against Eq. (4) with $c_k = 0$.

The analyses presented above, allow us to evaluate the regimes at which each friction component dominates. To that end, the upper envelope fits presented in Fig. 4, which assume negligible inner-surface contribution, are used to extrapolate the corner and edge friction components to mesoscale contact areas. This can be compared to the inner-surface contribution by performing separate periodic boundary conditions calculations, that eliminate edge and corner effects, and allow us to extract the inner surface contribution to the friction per unit area, $c_k$. Fig. 4d and 4h present the corresponding extrapolations demonstrating the shape-dependent cross-over from corner and edge to surface dominated friction. Specifically, for the kinetic friction, Eq. (4) can be used to evaluate the transition point between corner to edge dominated friction via the requirement that $a_k \approx b_k A^{\frac{1}{2}}$, yielding a side length of $l \approx a_k/b_k = 340, 139, 180, 60$ nm for triangular, square, rectangular, and hexagonal flakes, respectively. Similarly, the transition point to bulk dominated friction is obtained from the requirement that $b_k A^{\frac{1}{2}} = c_k A$ yielding a side length of $l \approx b_k/c_k$ that falls in the range $O(100$ nm$) - O(10$ µm$)$ depending on the shape of the flake. A similar analysis for static friction yields a transition from edge to surface dominated friction at a larger side length of $O(100$ µm$)$. These scaling transition points are larger than those previously predicted for circular flakes,[11] and are consistent with most experimental observations demonstrating edge dominated friction (sublinear scaling) in the $O(100$ nm$) - O(1$ µm$)$ side length range.[16-19,24,27-30]

**Effect of Interfacial Twist**

The results presented above are given for aligned heterogeneous contacts. To evaluate the effect of marginal twisting, which in the case of rigid sliding leads to the absence of static friction scaling with contact dimensions,[34] we performed sliding simulations of square and triangular flakes twisted by an angle of $\theta = 1°$ and sliding atop $h$-BN along the same scanline as that used for the aligned interfaces. In this configuration, the moiré superlattice of period 9.9 nm is rotated by 44.5° with respect to the $h$-BN lattice vectors.

Figure 5 presents the static (panels a and b) and kinetic (panels c and d) friction forces as a function of contact dimensions for square (panels a and c) and triangular (b and d) flakes. Both static and kinetic friction forces are significantly lower than those obtained for the aligned heterogeneous interface. We note that the higher kinetic friction observed for small square flakes and some triangular flakes results from sliding induced ABA⇌ABC corner stacking instability rather than elasticity effects (see SM Movie 4). Similar to the aligned case, kinetic friction is dominated by corner energy dissipation (see Fig. 6), where due to the sharper corner angle triangular flakes exhibit larger kinetic friction than their square counterparts (apart from the cases dominated by corner stacking instability) at the contact dimensions considered. Regarding the scaling, apart from some periodic variations no increase of static or kinetic friction with contact area is observed up to the largest system considered, which is consistent with rigid sliding model predictions.[34]

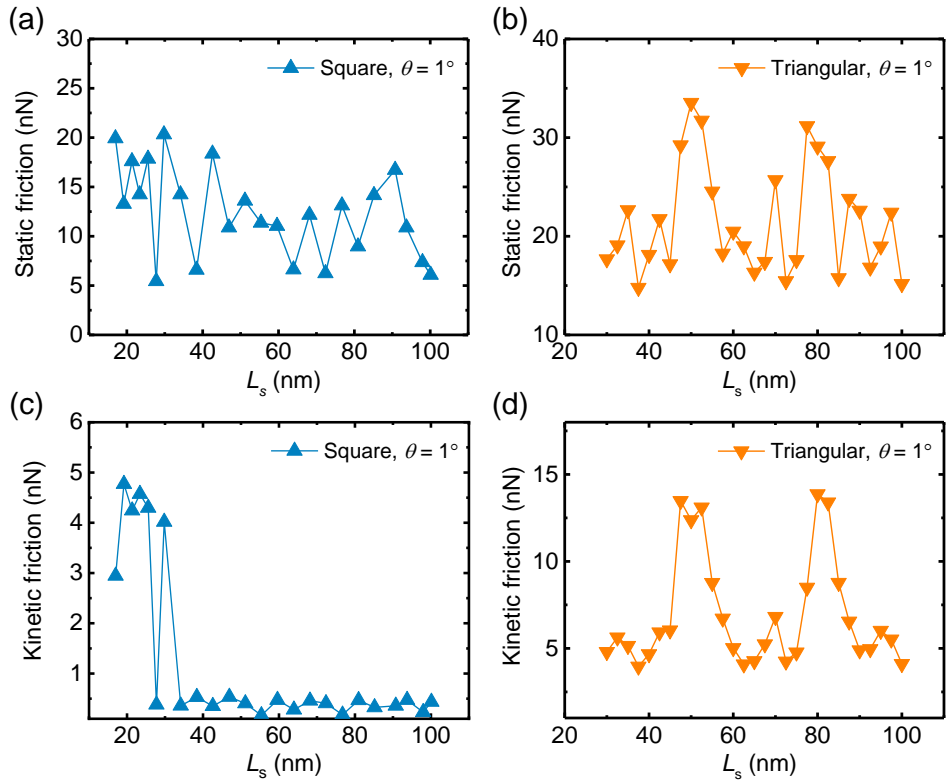

**Figure 5. Scaling of the static and kinetic friction forces with the dimensions of marginally twisted ($\theta = 1°$) square and triangular flakes.** Static friction force as a function of contact side length for heterogeneous graphene/$h$-BN square (a) and triangular (b) contacts. (c) and (d) same as (a) and (b) for the kinetic friction.

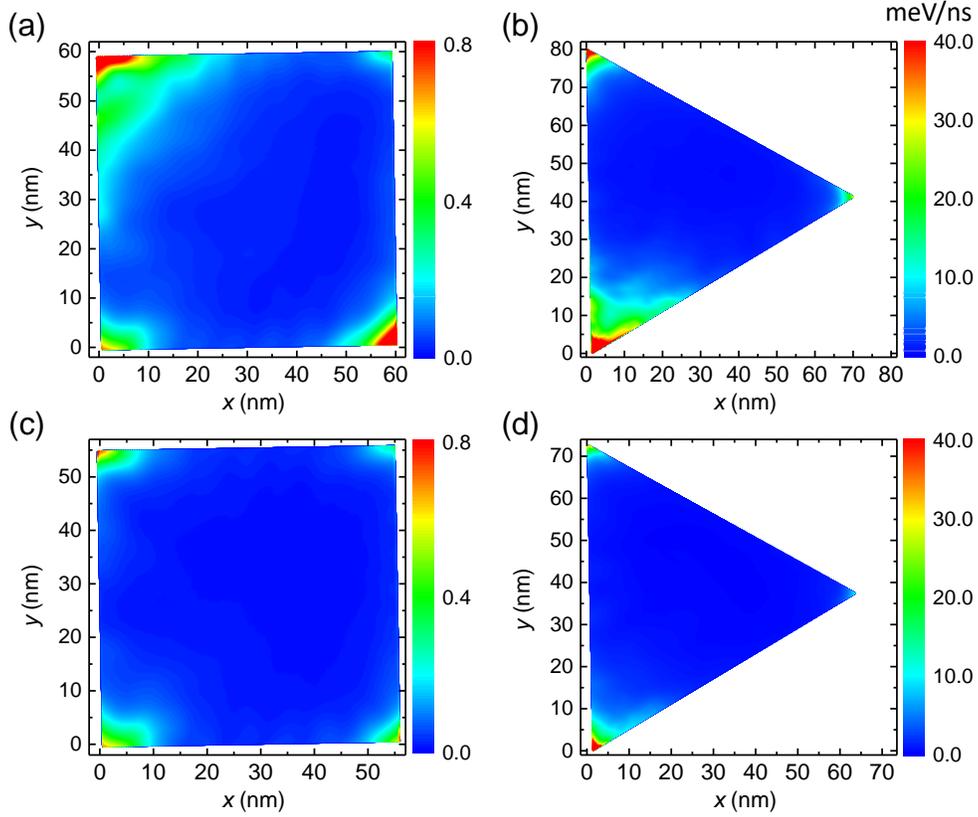

**Figure 6. Energy dissipation power distribution for marginally twisted ($\theta = 1°$) square and triangular flakes.** Energy dissipation maps of the middle layer of (a) square and (b) triangular flakes measured for system dimensions that correspond to local maxima in the kinetic friction with side length. (c) and (d) same as (a) and (b) for local minima in the kinetic friction. Atomic energy dissipation power is calculated by averaging over four sliding periods.

**Discussion and Concluding Remarks**

The results presented above demonstrate that elasticity effects can dominate kinetic friction of incommensurate layered material interfaces already at the nanometer scale. This extends previous theoretical predictions that elasticity effects in such systems should become important at a range of 10-100 µm.[7,25,30,46,47] While the latter prediction is based on shear induced lattice reconstruction (referred to as dislocations) that leads to local pinned surface region, our simulations reveal a different energy dissipation mechanism of incomplete moiré tile pinning at the finite slider corners and edges. Specifically, we find that the sharpness of the corner and the relative orientation of the sides with respect to the moiré superlattice and sliding direction dictate energy dissipation, thus paving the way for shape tailoring of friction. For aligned polygonal heterogeneous graphene/*h*-BN interfaces, friction is found to scale with side length, consistent with experimental observations.[17,30,48] Notably, marginal twisting leads to lack of such scaling, thus allowing to achieve microscale superlubricity.

**Acknowledgments**

X. G. acknowledges the postdoctoral fellowships of the Sackler Center for Computational Molecular and Materials Science and the Ratner Center for Single Molecule Science at Tel Aviv University. M. U. acknowledges the financial support of the BSF-NSF grant no. 2023614. O. H. is grateful for the generous financial support of the Heineman Chair in Physical Chemistry and Tel Aviv University Center for Nanoscience and Nanotechnology. W. O., and Z. L. would like to acknowledge supports from the National Natural Science Foundation of China (Nos. 12102307, 12172260 and 11890673), the Key Research and Development Program of Hubei Province (2021BAA192), the Natural Science Foundation of Hubei Province (2021CFB138), the Fundamental Research Funds for the Central Universities (2042023kf0233) and the startup fund of Wuhan University. Part of the simulations presented in this work were carried out at the National Supercomputer TianHe-1(A) Center in Tianjin.

**References**

1. Hod, O., Meyer, E., Zheng, Q. & Urbakh, M. Structural superlubricity and ultralow friction across the length scales. *Nature* **563**, 485-492 (2018).
2. Dienwiebel, M., Verhoeven, G. S., Pradeep, N., Frenken, J. W. M., Heimberg, J. A. & Zandbergen, H. W. Superlubricity of Graphite. *Phys. Rev. Lett.* **92**, 126101 (2004).
3. Liu, Z., Yang, J., Grey, F., Liu, J. Z., Liu, Y., Wang, Y., Yang, Y., Cheng, Y. & Zheng, Q. Observation of microscale superlubricity in graphite. *Phys. Rev. Lett.* **108**, 205503 (2012).
4. Filippov, A. E., Dienwiebel, M., Frenken, J. W. M., Klafter, J. & Urbakh, M. Torque and Twist against Superlubricity. *Phys. Rev. Lett.* **100**, 046102 (2008).
5. Leven, I., Krepel, D., Shemesh, O. & Hod, O. Robust Superlubricity in Graphene/*h*-BN Heterojunctions. *J. Phys. Chem. Lett.* **4**, 115-120 (2013).
6. Mandelli, D., Leven, I., Hod, O. & Urbakh, M. Sliding friction of graphene/hexagonal–boron nitride heterojunctions: a route to robust superlubricity. *Sci. Rep.* **7**, 10851 (2017).
7. Song, Y., Mandelli, D., Hod, O., Urbakh, M., Ma, M. & Zheng, Q. Robust microscale superlubricity in graphite/hexagonal boron nitride layered heterojunctions. *Nat. Mater.* **17**, 894 (2018).
8. Liao, M. *et al.* Ultra-low friction and edge-pinning effect in large-lattice-mismatch van der Waals heterostructures. *Nat. Mater.* **21**, 47-53 (2021).
9. Mandelli, D., Ouyang, W., Hod, O. & Urbakh, M. Negative Friction Coefficients in Superlubric Graphite-Hexagonal Boron Nitride Heterojunctions. *Phys. Rev. Lett.* **122**, 076102 (2019).
10. Wang, J., Khosravi, A., Vanossi, A. & Tosatti, E. Colloquium: Sliding and pinning in structurally lubric 2D material interfaces. *Rev. Mod. Phys.* **96**, 011002 (2024).
11. Wang, J., Cao, W., Song, Y., Qu, C., Zheng, Q. & Ma, M. Generalized Scaling Law of Structural Superlubricity. *Nano Lett.* **19**, 7735-7741 (2019).
12. Wang, J., Ma, M. & Tosatti, E. Kinetic friction of structurally superlubric 2D material interfaces. *J. Mech. Phys. Solids* **180**, 105396 (2023).
13. Gao, X., Ouyang, W., Hod, O. & Urbakh, M. Mechanisms of frictional energy dissipation at graphene grain boundaries. *Phys. Rev. B* **103**, 045418 (2021).
14. Gao, X., Ouyang, W., Urbakh, M. & Hod, O. Superlubric polycrystalline graphene interfaces. *Nat. Commun.* **12**, 5694 (2021).
15. Gao, X., Urbakh, M. & Hod, O. Stick-Slip Dynamics of Moiré Superstructures in Polycrystalline 2D Material Interfaces. *Phys. Rev. Lett.* **129**, 276101 (2022).
16. Hartmuth, F., Dietzel, D., de Wijn, A. S. & Schirmeisen, A. Friction vs. Area Scaling of Superlubric NaCl-Particles on Graphite. *Lubricants* **7** (2019).
17. Barabas, A. Z., Sequeira, I., Yang, Y., Barajas-Aguilar, A. H., Taniguchi, T., Watanabe, K. & Sanchez-Yamagishi, J. D. Mechanically reconfigurable van der Waals devices via low-friction gold sliding. *Sci. Adv.* **9**, eadf9558.
18. Dietzel, D., Brndiar, J., Štich, I. & Schirmeisen, A. Limitations of Structural Superlubricity: Chemical Bonds versus Contact Size. *ACS Nano* **11**, 7642-7647 (2017).
19. Qu, C., Wang, K., Wang, J., Gongyang, Y., Carpick, R. W., Urbakh, M. & Zheng, Q. Origin of Friction in Superlubric Graphite Contacts. *Phys. Rev. Lett.* **125**, 126102 (2020).
20. Liu, Y., Ren, J., Kong, D., Shan, G. & Dou, K. Edge-pinning effect of graphene nanoflakes sliding atop graphene. *Materials Today Physics* **38**, 101266 (2023).

21. Ying, P., Natan, A., Hod, O. & Urbakh, M. Effect of Interlayer Bonding on Superlubric Sliding of Graphene Contacts: A Machine-Learning Potential Study. *ACS Nano* **18**, 10133-10141 (2024).
22. Wang, K., He, Y., Cao, W., Wang, J., Qu, C., Chai, M., Liu, Y., Zheng, Q. & Ma, M. Structural superlubricity with a contaminant-rich interface. *J. Mech. Phys. Solids* **169**, 105063 (2022).
23. Müser, M. H., Wenning, L. & Robbins, M. O. Simple Microscopic Theory of Amontons's Laws for Static Friction. *Phys. Rev. Lett.* **86**, 1295-1298 (2001).
24. Dietzel, D., Ritter, C., Mönninghoff, T., Fuchs, H., Schirmeisen, A. & Schwarz, U. D. Frictional Duality Observed during Nanoparticle Sliding. *Phys. Rev. Lett.* **101**, 125505 (2008).
25. Sharp, T. A., Pastewka, L. & Robbins, M. O. Elasticity limits structural superlubricity in large contacts. *Phys. Rev. B* **93**, 121402 (2016).
26. Müser, M. H. Structural lubricity: Role of dimension and symmetry. *Europhysics Letters* **66**, 97 (2004).
27. Özoğul, A., İpek, S., Durgun, E. & Baykara, M. Z. Structural superlubricity of platinum on graphite under ambient conditions: The effects of chemistry and geometry. *Appl. Phys. Lett.* **111**, 211602 (2017).
28. Cihan, E., İpek, S., Durgun, E. & Baykara, M. Z. Structural lubricity under ambient conditions. *Nat. Commun.* **7**, 12055 (2016).
29. Dietzel, D., Feldmann, M., Schwarz, U. D., Fuchs, H. & Schirmeisen, A. Scaling Laws of Structural Lubricity. *Phys. Rev. Lett.* **111**, 235502 (2013).
30. Koren, E., Lörtscher, E., Rawlings, C., Knoll, A. W. & Duerig, U. Adhesion and friction in mesoscopic graphite contacts. *Science* **348**, 679-683 (2015).
31. Li, H., Wang, J., Gao, S., Chen, Q., Peng, L., Liu, K. & Wei, X. Superlubricity between $MoS_2$ Monolayers. *Adv. Mater.* **29**, 1701474 (2017).
32. Koren, E. & Duerig, U. Moiré scaling of the sliding force in twisted bilayer graphene. *Phys. Rev. B* **94**, 045401 (2016).
33. de Wijn, A. S. (In)commensurability, scaling, and multiplicity of friction in nanocrystals and application to gold nanocrystals on graphite. *Phys. Rev. B* **86**, 085429 (2012).
34. Yan, W., Gao, X., Ouyang, W., Liu, Z., Hod, O. & Urbakh, M. Shape-dependent friction scaling laws in twisted layered material interfaces. *J. Mech. Phys. Solids* **185**, 105555 (2024).
35. Brenner, D. W., Shenderova, O. A., Harrison, J. A., Stuart, S. J., Ni, B. & Sinnott, S. B. A second-generation reactive empirical bond order (REBO) potential energy expression for hydrocarbons. *J. Phys.: Condens. Matter* **14**, 783-802 (2002).
36. Tersoff, J. New empirical approach for the structure and energy of covalent systems. *Phys. Rev. B* **37**, 6991-7000 (1988).
37. Kolmogorov, A. N. & Crespi, V. H. Registry-dependent interlayer potential for graphitic systems. *Phys. Rev. B* **71**, 235415 (2005).
38. Leven, I., Maaravi, T., Azuri, I., Kronik, L. & Hod, O. Interlayer Potential for Graphene/*h*-BN Heterostructures. *J. Chem. Theory Comput.* **12**, 2896 (2016).
39. Leven, I., Azuri, I., Kronik, L. & Hod, O. Inter-Layer Potential for Hexagonal Boron Nitride. *J. Chem. Phys.* **140**, 104106 (2014).

40  Maaravi, T., Leven, I., Azuri, I., Kronik, L. & Hod, O. Interlayer Potential for Homogeneous Graphene and Hexagonal Boron Nitride Systems: Reparametrization for Many-Body Dispersion Effects. *J. Phys. Chem. C* **121**, 22826 (2017).
41  Ouyang, W., Mandelli, D., Urbakh, M. & Hod, O. Nanoserpents: Graphene Nanoribbon Motion on Two-Dimensional Hexagonal Materials. *Nano Lett.* **18**, 6009 (2018).
42  Ouyang, W., Azuri, I., Mandelli, D., Tkatchenko, A., Kronik, L., Urbakh, M. & Hod, O. Mechanical and Tribological Properties of Layered Materials under High Pressure: Assessing the Importance of Many-Body Dispersion Effects. *J. Chem. Theory Comput.* **16**, 666 (2020).
43  Plimpton, S. Fast Parallel Algorithms for Short-Range Molecular Dynamics. *J. Comput. Phys.* **117**, 1-19 (1995).
44  Vanossi, A., Manini, N. & Tosatti, E. Static and dynamic friction in sliding colloidal monolayers. *Proc. Natl. Acad. Sci. U.S.A* **109**, 16429-16433 (2012).
45  Ouyang, W., Sofer, R., Gao, X., Hermann, J., Tkatchenko, A., Kronik, L., Urbakh, M. & Hod, O. Anisotropic Interlayer Force Field for Transition Metal Dichalcogenides: The Case of Molybdenum Disulfide. *J. Chem. Theory Comput.* **17**, 7237-7245 (2021).
46  Shen, Y. & Wu, H. Interlayer shear effect on multilayer graphene subjected to bending. *Appl. Phys. Lett.* **100** (2012).
47  Bosak, A., Krisch, M., Mohr, M., Maultzsch, J. & Thomsen, C. Elasticity of single-crystalline graphite: Inelastic x-ray scattering study. *Phys. Rev. B* **75**, 153408 (2007).
48  Li, Y., He, W., He, Q.-C. & Wang, W. Contributions of Edge and Internal Atoms to the Friction of Two-Dimensional Heterojunctions. *Phys. Rev. Lett.* **133**, 126202 (2024).